\documentclass[11pt]{article}

\pdfoutput=1
\usepackage[pdftex]{color}
\usepackage{amssymb}
\usepackage{amsthm}
\usepackage{amsmath}
\usepackage{latexsym}
\usepackage{amscd}
\usepackage{graphicx}
\usepackage[pdftex, colorlinks=true, citecolor=green]{hyperref}
\usepackage{lscape}
\usepackage{setspace}
\usepackage{multirow}
\usepackage{enumerate}

\newcommand{\ds}{\displaystyle}

\newcommand{\ben}{\begin{equation}}     
\newcommand{\eeqn}{\end{equation}}
\newcommand{\bey}{\begin{eqnarray}}
\newcommand{\eey}{\end{eqnarray}}

\begin{document}

\noindent {\Large
\textbf{An energy balance model of carbon's effect on climate change}
}
\\\\
Lucas Benney$^{1}$, Anca R\v{a}dulescu$^{2,*}\footnotetext{*Corresponding author: Assistant Professor, Department of Mathematics, State University of New York at New Paltz; New York, USA; Phone: (845) 257-3532; Email: radulesa@newpaltz.edu}$
\\
\indent $^1$ Department of Mathematics, SUNY New Paltz, NY 12561
\\
\indent $^2$ Department of Engineering,  SUNY New Paltz, NY 12561

\vspace{7mm}
\begin{abstract}

 Due to climate change, the interest of studying our climatic system using mathematical modeling has become tremendous in recent
years. One well-known model is Budyko's system, which represents the coupled evolution of two variables, the ice-line and the
average earth surface temperature. The system depends on natural parameters, such as the earth albedo, and the amount A of carbon in the atmosphere.

We introduce a 3-dimensional extension of this model in which we regard A as the third coupled variable of the system. We analyze
the phase space and dependence on parameters, looking for Hopf bifurcations and the birth of cycling behavior. We interpret the cycles as climatic oscillations triggered by the sensitivity in our regulation of carbon emissions at extreme temperatures

\end{abstract}

\section{Introduction}

\begin{figure}[h!]
\begin{center}
\includegraphics[width=0.8\textwidth]{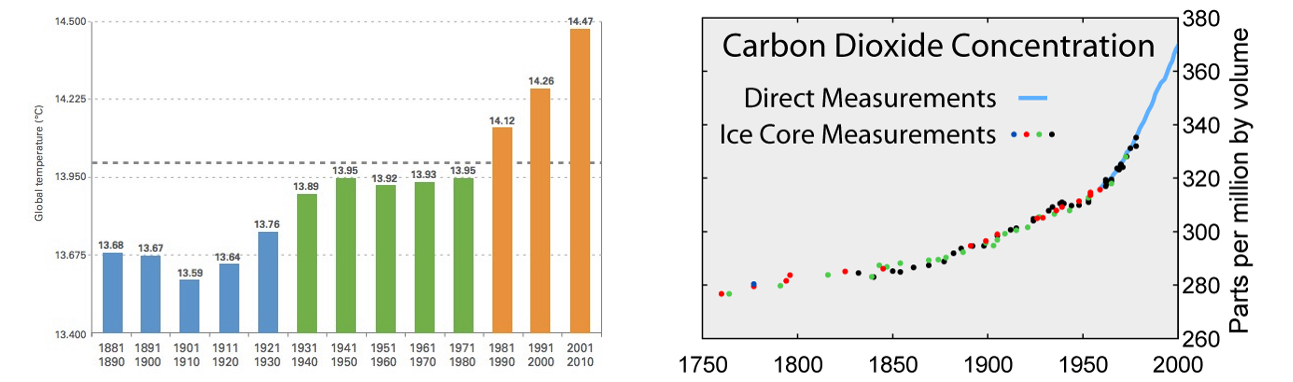}
\end{center}
\label{temp_carb}
\caption{\emph{{\bf Earth's average temperature and Carbon Dioxide concentration annually.} {\bf Left.} Earth's average surface temperature over the past century. {\bf Right.} Measured levels of Carbon Dioxide concentrations annually.}}
\label{temp_increase}
\end{figure}

Earth's climate is constantly changing. This change has become much more prominent over the past century, during which time the average surface temperature of the Earth has increased by almost a full degree Celsius. Figure~\ref{temp_increase} shows side by side the average annual surface temperature of the Earth (on the left), and the annual concentration of Carbon Dioxide (on the right). Both graphs show a similar,  almost exponential increase with time. The correlation between the two trends suggests that surface temperature is related to carbon dioxide levels. 

Over the past hundreds of millions of years the earth's average surface temperature has gone through many different cycles. There have been times where it was so warm that reptiles could survive above the arctic circle and times when glaciers covered most of the earth. The average surface temperature of the earth has been rising consistently over the past hundred years. We believe that the temperatures will continue to rise, and some big questions relate to this: How long until the cycle reaches a maximum temperature value? Will the earth be inhabitable by humans at this maximum? Can our behavior, such as green house gas emissions, or excess use of fossil fuels, change the established cycles?

\subsection{The Budyko model}

 Russian Climatologist Mikhail Budyko (July 28, 1920 - December 10, 2001) is known for being one of the founders of physical climatology. He formulated the integro-differential equation~\eqref{Budyko} by assuming Earth's annual average surface temperature at a given latitude $T = T(t,y)$ (in $^o C$) can be represented as a function of time (t) and the sine of the latitude $ (y = \sin(\theta)) $.

\begin{equation}
R\frac{\partial T}{\partial t} = Qs(y)(1-\alpha(y,\eta)) - (A+BT)+C(\overline{T}-T)
\label{Budyko}
\end{equation}

\noindent where 

$$\overline{T}(t) = \int_0^1 T(t,y)  \; dy$$

This energy balance model (EBM), referred to as Budyko's model~\cite{mcgehee2014quadratic}, is based on simplifications of factors that cause the Earth's surface temperature to change. The temperature is assumed to be constant on a given latitude circle and symmetric across the equator (hence $y$ ranges from 0 at the equator to 1 at the north pole). It is assumed that the average temperature is a continuous function with respect to time. $\overline{T}$ is the global annual mean temperature.

Both sides of the equation are expressed in $W/m^2$ (since the parameters $R$, representing the average heat capacity of the earth's surface, is measured in $J/m^2$ $^\circ C$, and the temperature $T$ is measures in $^\circ C$. Parameter values and their corresponding units are shown in Table~\ref{params}.

A key component of the equation is $\alpha(y,\eta)$, which represents the albedo  (amount of shortwave solar radiation reflected back into space) for the earth, defined in Equation~\eqref{albedo}:

\begin{equation}
\alpha(y,\eta) = \left\{
\begin{array}{lr}
 \alpha_w,&\text{if } y < \eta,\\
 \alpha_s,&\text{if } y > \eta
\end{array}
\right.
\label{albedo}
\end{equation}

The albedo function is defined piecewise and depends on the latitude $y$ and on the position $\eta$ of the iceline (the glacier line), so that for $y > \eta$  the earth is covered in ice, with an albedo of $\alpha_s = 0.62$, and for $y < \eta$ the Earth is ice-free, with  $\alpha_w = 0.32$. More radiation is reflected for ice than water, producing a larger albedo $\alpha$ for above-iceline latitudes.

The parameter $Q$ represents the ``solar constant,'' or the annual global mean insolation. The distribution of that insolation over latitude is represented by $s(y)$, with $\int_0^1 s(y) \; dy = 1$. Putting these pieces together, one can interpret the first term on the right to represent the amount of incoming solar energy absorbed by the Earth. 

Outgoing longwave radiation (OLR) is approximated by the second term on the right $A+BT$ (the values of $A$ and $B$, determined by satellite measurements, are shown in Table~\ref{params}). This term can be looked at as a measure of the amount of Carbon (or more generally, greenhouse gases) in the atmosphere. If there is more carbon in the atmosphere, less radiation escapes, therefore this term will be smaller. 

The final term on the right represents convection, i.e., heat is transfer between latitudes. In Budyko's model it is assumed that over the period of a year this transfer of heat can be represented as the difference between the average global surface temperature $\overline{T}$ and the surface temperature at the current latitude, then multiplying this difference by a proportionality constant $(C)$ (also determined by satellites).

\begin{table}[h!]
\begin{center}
\begin{tabular}{|p{3cm}|p{3cm}|p{3cm}|}
\hline
{\bf Parameter} & {\bf Value} & {\bf Units} \\
\hline \hline
$Q$ & 343 & W/m$^2$ \\
\hline
$s_2$ & -0.482 & Dimensionless\\
\hline
$A$ & 202 & W/m$^2$ \\
\hline
$B$ & 1.9 & W/m$^2$ $^\circ C$ \\
\hline
$D$ & 3.04 & W/m$^2$ $^\circ C$\\
\hline
$R$ & $4 \times 10^9$ & J/m$^2$ $^\circ C$ \\
\hline
$\alpha_1$ & 0.32 & Dimensionless \\
\hline
$\alpha_2$ & 0.62 & Dimensionless\\
\hline
$T_c$ & -10 & $^\circ C$ \\
\hline
$\Omega$ & $1.5  \cdot 10^{11}$ & $J/m^2$ \\
\hline
$\varepsilon$ & $3.9 \cdot 10^{-13}$ &  \\
\hline
\end{tabular}
\end{center}
\caption{\emph{{\bf Model parameter values and units}, as per the original references.}}
\label{params}
\end{table}

\subsection{Iceline dynamics}
\label{iceline}

More recent work by McGehee and Widiasih incorporated temperature-triggered iceline variability in the original EBM, and formulated the temperature-iceline dynamics as a system of two coupled differential equations~\cite{mcgehee2012simplification,mcgehee2014quadratic}. Below, we only briefly outline the derivation of the coupled equations; a comprehensive explanation can be found in the original reference.

First, as simpler approximation was introduced for the distribution of insolation $s(y)$, which was expressed in terms of the angle $\beta$ between the earth's axis of rotation and perpendicular to plane of earth's orbit~\cite{mcgehee2012paleoclimate}:

\begin{equation}
s(y,\beta) = \frac{2}{\pi^2}\int\limits_{0}^{2\pi} \sqrt{1 - \left( \sqrt{1 - y^2}\sin(\beta)\cos(\theta) - y\cos(\beta) \right)^2} \; d\theta
\end{equation}

\noindent  A quadratic approximation (to within $2\%$ error) led to and expression for $s(y)$ of the form:

\begin{equation}
s(y) \approx 1 + s_2p_2(y)
\end{equation}

\noindent where 

\begin{equation}
p_2(y) = \frac{1}{2} \left( 3y^2-1 \right)
\end{equation}

\noindent It was shown~\cite{mcgehee2012paleoclimate} that $s_2 \approx - 0.482$. This approximation allowed for simpler methods of solving for equilibrium solutions to Budyko's mode for a fixed iceline $\eta$, without taking a large toll on accuracy. The discontinuity of the albedo function $\alpha(y,\eta)$ at $y=\eta$ leads to the question of defining $T_\eta^*(y)$ at $y=\eta$. This was taken to be the average of the two side limits: 

$$T_\eta^*(\eta) = \frac{T_\eta^*(\eta-) + T_\eta^*(\eta+)}{2}$$

\noindent leading to the equilibria:

\begin{displaymath}
T_\eta^*(y) = \left\{
\begin{array}{lr}
\ds \frac{1}{B+C} \left( Qs(y)(1-\alpha_w) - A + C\overline{T_\eta^*} \right), & \text{ if } y < \eta,\\\\
\ds \frac{1}{B+C} \left( Qs(y)(1-\alpha_s) - A + C\overline{T_\eta^*} \right), & \text{ if } y > \eta,\\\\
\ds \frac{1}{B+C} \left( Qs(\eta) \left( 1-\frac{\alpha_w + \alpha_s}{2} \right) - A + C\overline{T_\eta^*} \right), & \text{ if } y = \eta
\end{array}
\right.
\end{displaymath}

\noindent where

\begin{equation}
\overline{T_\eta^*} = \frac{1}{B} \left( Q \left[ 1-\alpha_s - (\alpha_s - \alpha_w) \left( \eta + \frac{s_2}{2}(\eta^3 - \eta) \right) \right] - A \right)
\end{equation}

It has been observed that glaciers form at a temperature of approximately  $-10^\circ C$. So if it is assumed that the ice line is stationary, the average temperature across that iceline should be a critical temperature, $T_c = -10^\circ C$. This means that, with the assumption of a fixed iceline, $T_\eta^*(y)$. However, in light of observations of iceline dynamics over many years, the stationarity assumption is not realistic~\cite{mcgehee2012paleoclimate}.

That is because if the equilibrium temperature profile is greater than the critical temperature, the ice line will retreat towards $\eta = 1$ and the opposite if the temperature is below $T_c$. According to this considerations, Widiasih~\cite{widiasih2013dynamics} introduced the rate of change of the ice line latitude as being proportional to the difference between the equilibrium temperature profile and the critical temperature (with proportionality constant $\varepsilon$ extremely small, since it represents the time scale of glacier dynamics):

\begin{equation}
\frac{d\eta}{dt} = \varepsilon(T_\eta^*(\eta) - T_c)
\label{iceline_eq}
\end{equation}

The energy required to melt ice was also included in the original EBM, via a parameter $\Omega$, representing the amount of energy required to melt a square meter of ice:

\begin{equation}
\Omega = 1.5  \cdot 10^{11} J/m^2
\end{equation}

With these extensions, the Budyko model with iceline dynamics becomes a system of coupled equations in $(T,\eta)$:

\begin{eqnarray}
\frac{d\eta}{dt} &=& \varepsilon(T_\eta^*(\eta) - T_c) \nonumber \\
R\frac{\partial T}{\partial t} &=& Qs(y)(1-\alpha(y,\eta)) - (A+BT)+C(\overline{T}-T) - \Omega\frac{d\eta}{dt}
\end{eqnarray}

 In the original reference~\cite{mcgehee2014quadratic}, the authors use a Legendre expansion and change of variables to find a two dimensional invariant subspace that yields to a simple system of two coupled differential equations. For the remainder of this section, we only sketch the main steps of this process. 

Assuming that $T_\eta^*(y)$ has piecewise quadratic equilibrium solutions, it can be expressed as.

\begin{displaymath}
T_\eta^*(y) = \left\{
\begin{array}{lr}
U(y)&: y < \eta,\\
V(y)&: y > \eta,\\
\frac{1}{2}(U(\eta) + V(\eta))&: y = \eta,
\end{array}
\right.
\end{displaymath}

\noindent where both $U$ and $V$ are assumed to be quadratic on the interval [0,1]. The system can be then written in terms of $U$ and $V$ as:

\begin{eqnarray}
U(y,t) &=& u_0(t)p_0(y) + u_2(t)p_2(y), \nonumber \\
V(y,t) &=& v_0(t)p_0(y) + v_2(t)p_2(y),\\
s(y) &=& s_0p_0(y) + s_2p_2(y). \nonumber
\end{eqnarray}

Since $s$ is an even function, $U$ and $V$ must also be even. For convenience, $p_0(y)$ and $p_2(y)$ were taken to be the first two even Legendre polynomials. 

\begin{eqnarray}
p_0(y) &=& 1, \nonumber \\
p_2(y) &=& \frac{1}{2}(3y^2 - 1),
\end{eqnarray}

\noindent (recall that, since $s(y)$ was normalized, $s_0 = 1$). Introducing back in the original equation:

\begin{eqnarray}
\frac{d\eta}{dt} &=& \varepsilon(T_\eta^*(\eta) - T_c), \nonumber \\
R\frac{du_0}{dt} &=& Q(1 - \alpha_w) - A - (B+C)u_0 + C\overline{T}(\eta) - \varepsilon\Omega(T_\eta^* - T_c), \nonumber \\
R\frac{dv_0}{dt} &=&  Q(1 - \alpha_s) - A - (B+C)v_0 + C\overline{T}(\eta) - \varepsilon\Omega(T_\eta^* - T_c),\\
R\frac{du_2}{dt}&=& Qs_2(1 - \alpha_w) - (B+C)u_2, \nonumber\\
R\frac{dv_2}{dt}&=& Qs_2(1 - \alpha_s) - (B+C)v_2, \nonumber
\end{eqnarray}

After a change of variables, the system becomes more manageable and simplifies to:

\begin{eqnarray}
\frac{d\eta}{dt} &=& \varepsilon(T_\eta^*(\eta) - T_c), \nonumber \\
R\frac{dw}{dt} &=& Q(1 - \frac{1}{2}(\alpha_w + \alpha_s)) - A - (B+C)w + C\overline{T}(\eta) - \varepsilon\Omega(T_\eta^* - T_c), \nonumber \\
R\frac{dz}{dt} &=& Q(\alpha_s - \alpha_w) - (B+C)z, \\
R\frac{du_2}{dt} &=& Qs_2(1 - \alpha_w) - (B+C)u_2, \nonumber \\
R\frac{dv_2}{dt} &=& Qs_2(1 - \alpha_s) - (B+C)v_2, \nonumber
\end{eqnarray}
where
\begin{equation}
w = \frac{u_0 + v_0}{2}, 	z = u_0 - v_0. \nonumber
\end{equation}

If we examine the equations for $u_2,$ $v_2,$ $and$ $z$ we notice that they are linear differential equations that are decoupled from other variables. This allows us to conclude that there is a two dimensional invariant subspace defined by
\begin{eqnarray}
z &=& \frac{Q(\alpha_s - \alpha_w)}{B + C}, \nonumber \\
u_2 &=& \frac{Qs_2(1 - \alpha_w)}{B + C}, \\
v_2 &=& \frac{Qs_2(1 - \alpha_s)}{B + C}.\nonumber
\end{eqnarray}

On this subspace, the two-dimensional system becomes:

\begin{eqnarray}
 \frac{d\eta}{dt} &=& \varepsilon \left[ w + \frac{Qs_2}{2} \left( 1 - \frac{\alpha_w + \alpha_s}{2} \right) \left( \frac{3\eta^2 - 1}{B+C} \right) - T_c \right]  \\
R\frac{dw}{dt} &=& Q \left( 1 - \frac{\alpha_w +\alpha_s}{2} \right) - A + \frac{QC}{B + C}(\alpha_s - \alpha_w) \left( \eta - \frac{1}{2} + \frac{s_2}{2}(\eta^3 - \eta) \right) - Bw - \Omega\frac{d \eta}{d t}. \nonumber
\label{iceline_sys}
\end{eqnarray}


\subsection{Nonlinear dynamics}

These two coupled equations define a two-dimensional, continuous time, nonlinear dynamical system. A typical analysis of such as system begins with finding nullclines and equilibria, with a linear analysis around equilibria to establish local stability, and may continue with a search for global stability features, with searching for limit cycles and with an analysis of the system's dependence on parameters. Often times, however, even seemingly simple nonlinear systems may be too complicated to solve directly, and this plan of action may fail, due to the difficulty of solving nonlinear algebraic equations, right at the first step: that of locating equilibria. In this case, numerical methods may offer adequate support (although numerical algorithms themselves may be problematic in the context of certain nonlinear features).

\begin{figure}[h!]
\begin{center}
\includegraphics[width=0.5\textwidth]{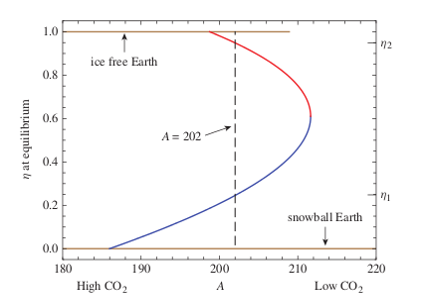}
\end{center}
\label{temp_greenhouse}
\caption{\emph{{\bf Saddle node bifurcation with respect to $A$.} The red plot represents a stable equilibrium curve w.r.t. $A$, and the blue curve an unstable (saddle) equilibrium. They collide at the bifurcation point $A=212$~\cite{walsh2013modeling}.}}
\label{A_bifurcation}
\end{figure}

The two-dimensional system~\eqref{iceline_sys} was approached with a combination of analytical and numerical computations to find the system's two equilibria, estimate their position and determine their stability (one was found to be a node with $\eta\sim 0.95$, and one a saddle with $\eta=0.25$). The authors also study the sensitivity of the system with respect to the parameter $A$ (see Figure~\ref{A_bifurcation}), to find a saddle node \emph{bifurcation}. A bifurcation is a state of the system that represents a sharp transition from one dynamic behavior to another, which may occur with changes in the number of equilibria and their stability, of with the formation or dissolution of a limit cycle (in Section~\ref{numerics}, we will show more examples of different types of bifurcations).

The bifurcation diagram of the system~\eqref{iceline_sys} with respect to $A$ (shown in Figure~\ref{A_bifurcation}) describes a sharp transition at a tipping point of $A \sim 212$. For values of $A$ smaller than $A ~\sim 198$, the system evolves away from the only unstable equilibrium in the natural range, and converges to an iceline $\eta=1$ (ice free Earth). For values of the parameter larger than $A \sim 212$, the system has no equilibria, and the system is driven to $\eta=0$ (snowball Earth). For intermediate values of $A$, the system will settle to a high iceline, and a livable average surface temperature of about 5 $^\circ C$. This implies, as expected, that excessive CO$_2$, reflected in low values of $A$, leads to an excessively hot Earth, but also suggests that too low values of CO$_2$, reflected in high $A$, lead to an equally undesirable outcome (frozen Earth). The existence of a range of CO$_2$ (and subsequently $A$) producing optimal temperatures for sustaining life on Earth motivated our interest in studying the coupling between CO$_2$ levels and surface temperature, and the effects of this coupling on the system dynamics.

 We worked on conceptually extending the original results in~\cite{mcgehee2012simplification} by introducing $A$ as a third variable in the system, then we investigated the nonlinear behavior of this extension. We focused in particular on the system's dependence on parameters, and on establishing whether is has any codimension one bifurcations (i.e., sharp transitions in dynamics that appear when only varying one parameter of the system). We looked in particular for saddle-node, Hopf and fold bifurcations (see~\cite{kuznetsov2013elements} for comprehensive definitions):
 
\vspace{2mm}
\noindent A \emph{saddle node} (or \emph{limit point}) bifurcation is a local bifurcation where two equilibria with different stability (typically a saddle and a node) collide and disappear (see Figure~\ref{A_bifurcation}). 

\vspace{2mm}
\noindent A \emph{Hopf bifurcation} is a local bifurcation where an equilibrium point changes stability, with the birth of a limit cycle. This happens if an eigenvalue $\lambda$ of the system's Jacobian matrix around the fixed point traverses the imaginary axis, i,e., changes from having $Re(\lambda)>0$ to a $Re(\lambda)<0$, with the bifurcation occurring at $Re(\lambda)=0$. A Hopf bifurcation can be supercritical (with a stable spiral equilibrium changing stability with the birth of a stable cycle) or subcritical (with an unstable equilibrium changing stability with the birth of an unstable cycle). One can establish the type of a Hopf bifurcation by considering the quadratic nonlinear terms of the system and compute the Lyapunov coefficient $\sigma$, which is $\sigma>0$ at a supercritical Hopf, and $\sigma<0$ at a subcritical Hopf.

\vspace{2mm}
\noindent A \emph{fold} (or \emph{limit point cycle}) bifurcation is a local bifurcation where two limit cycles with different stability collide and disappear (i.e., a limit point for cycles).\\

\noindent Since the direct computation of all the dynamic invariants that would permit classifications of these bifurcations would be practically intractable, we used in our analysis the Matcont software package (see Section~\ref{numerics}). Matcont~\cite{dhooge2003matcont} uses numerical continuation algorithms to track the changes in the behavior of the system as parameters are changed.


\subsection{Introducing dependence of carbon on temperature}
\label{A_param}

To begin constructing our extension of the classical model~\eqref{iceline_sys}, we start with the assumption that accumulation of greenhouse gases has a strong effect on the Earth OLR. Indeed, CO$_2$ absorbs energies with a wavelength of around 15 micrometers very easily. This happens to be included in the infrared region of the spectrum of light which has wavelengths ranging from 700 nanometers - 1 millimeter. This range of wavelengths is within the OLR wavelengths assumed in Budkyo's model (\url{www.ces.fau.edu/nasa/module-2/how-greenhouse-effect-works.php}). CO$_2$ is circulated throughout our atmosphere in many ways. The main four processes are: Decomposition of organic materials, respiration, photosynthesis, and combustion. Other than planting trees (or breathing less) the only control we humans have on the levels of CO$_2$ in the atmosphere is combustion (which we are clearly overdoing as a species).

There are, of course, other forms of greenhouse gases that effect OLR, for example: Methane (CH$_4$), Nitrous oxide (NO$_2$), and Ozone (O$_3$). In comparison to CO$_2$ all three listed greenhouse gases absorb more infrared radiation per molecule. So what motivates our specific focus on CO$_2$?

\begin{figure}[h!]
\begin{center}
\includegraphics[width=0.6\textwidth]{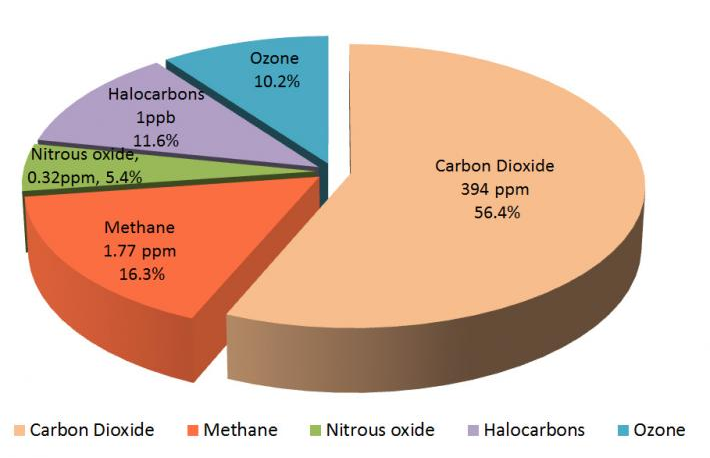}
\end{center}
\label{temp_greenhouse}
\caption{\emph{{\bf Percentages of greenhouse gases in the atmosphere.} We notice that Carbon Dioxide is the most abundant greenhouse gas. Figure obtained from NASA public domain:} \url{www.ces.fau.edu/nasa/module-2/how-greenhouse-effect-works.php}}
\label{greenhouse}
\end{figure}

Figure~\ref{greenhouse} shows the percentage of influence each greenhouse gas has on Earth's climate system. Although $CO_2$ is only present at a level of approximately 400 ppm (parts per million), small changes in the quantity of the greenhouse gas can have enormous effect on the earth's climate.

Recall that in our model the term $A + BT$, representing outgoing longwave radiation (ORL), depends on the level of greenhouse gases, which clearly affect ORL. Observing that $CO_2$ is by far the dominant greenhouse gas, we regard $A$ as a measure of the levels of $CO_2$ in the atmosphere, with lower values of $A$ representing high atmospheric levels of CO$_2$, and high values of $A$ making low CO$_2$.

Recall that Figure~\ref{temp_increase} shows how the levels of CO$_2$ and the temperature have changed over the past few centuries. Since $A$ reflects the levels of CO$_2$, then $A$ will also be time dependent, hence we will introduce $A(t)$ as a new (third coupled variable in our dynamic model. Moreover, as supported by the correlation between the graphs in the two figure panels, we suggest that $A$ depends implicitly on time by depending directly on the surface temperature. Indeed, a variety of studies have been quantifying with increasing success over the years the dependence of the Carbon footprint on temperature~\cite{trumbore1996rapid}. 

For example, there are geography-specific seasonal oscillations in levels of CO$_2$, believed to be based both on human activity and the life cycles of vegetation~\cite{randerson1999increases}. In the geographic zones where winter exists, photosynthesis slows during the winter months, causing less CO$_2$ to be removed from the atmosphere, followed by a fast increase in spring and summer, when the vegetation recovers and restarts photosynthesizing. In fall, large portions of the vegetation die and, through the decay process, emit large amounts of CO$_2$ back into the atmosphere. Humans, in turn, affect CO$_2$ levels by emitting large amounts of greenhouse gases (through maintenance needs like heating or cooling, transportation, industrial and agricultural activities). These also fluctuate throughout the year.

For our model, however, we are not primarily interested  in studying annual patterns in the the CO$_2$ levels at in specific geographic zones, but rather more global and longer-term effects produced by the coupling between temperature ($w(t)$, in our model) and CO$_2$ ($A(t)$, in our model). While a precise quantitative dependence of $A$ on $w$ has not yet been established, and would require considerable effort to express accurately, the literature in the field provides strong evidence of specific qualitative trends. For example, it was noted that the Carbon footprint differs by country, which is probably primarily consequential to climatic factors and resource availability. Recent climatology studies have suggested accounting for geography and climate in ranking countries' CO$_2$ emissions~\cite{parmesan2000impacts,neumayer2004national}. Cooling in the hot summer months raises significantly the consumption of electricity, which is the primary cause of raising CO$_2$, as does heating during cold winters (some believe that the excessive cold during recent winters is responsible for some of the 2013/4 raise in CO$_2$ emissions reported by the US Energy Infomation Administration). The relationship between climate and Carbon emissions is clearly mediated by the human response and resource availability, but other factors have been discussed, such as the influences on climate on the Earth's biota, which in turn contributes to driving CO$_2$ levels~\cite{parmesan2000impacts}.

Incorporating these ideas in out model, we specifically assume the following: Within an ideal range of temperatures (centered around $w_0=20^\circ C$, live organisms require less energy for artificial maintenance, and rather contribute themselves to maintaining function of the whole system in an optimal range. Plants are striving and efficiently remove atmospheric CO$_2$ through photosynthesis; people use reduced energy for heating and cooling, altogether increasing $A$ at intermediate values, by keeping $dA/dt$ positive. Deviations from the optimal range for $w$ lower the values of $A$: both excessively high and very low temperatures will diminish plant function and subsequently CO$_2$ removal, and will increase human Carbon footprint, with an over all negative differential between these two effects that leads to high CO$_2$ levels, hence switching $dA/dt$ to negative values and decreasing $A$. These effects exacerbate with larger $\lvert w-w_0 \rvert$, up to the point where life becomes unsustainable. For a start, we will assume that, past this threshold, plants slowly die out, and eventually human activity will also extinguish, so that, with no positive or negative contributions to the CO$_2$ levels, $dA/dt$ will return asymptotically to zero.

\begin{figure}[h!]
\begin{center}
\includegraphics[width=0.48\textwidth]{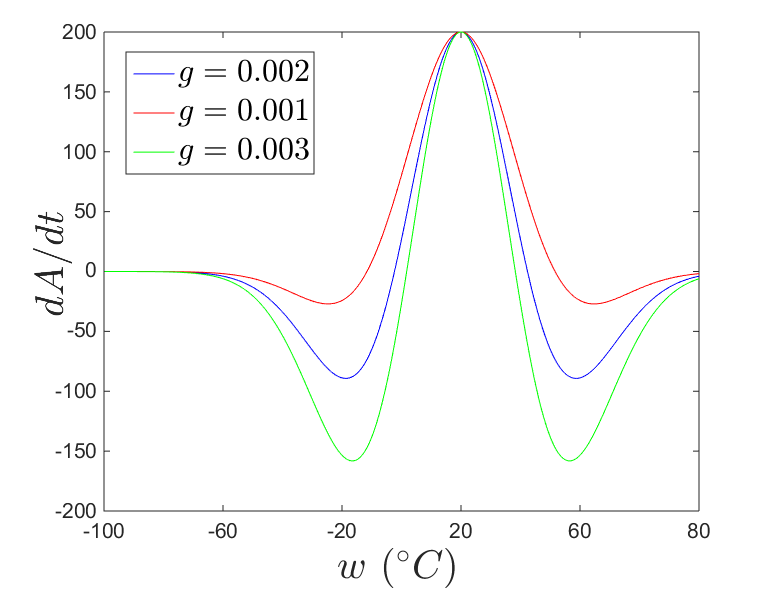}
\includegraphics[width=0.48\textwidth]{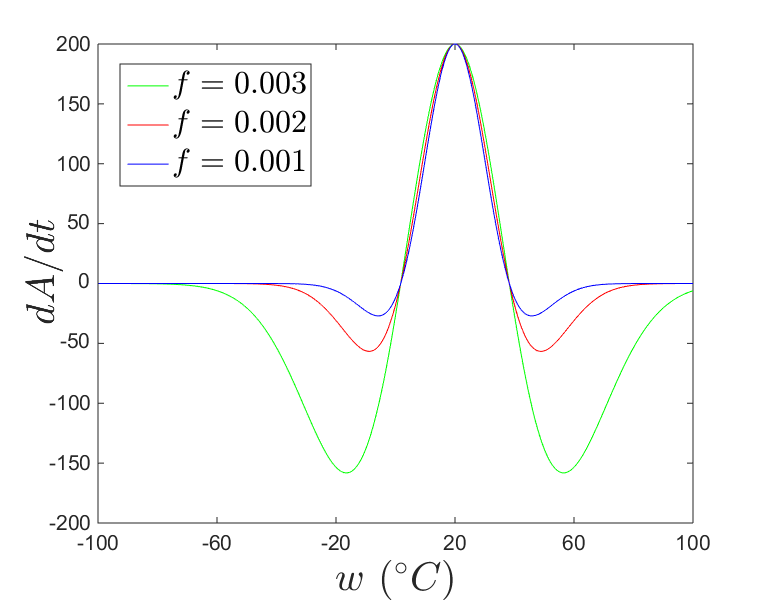}
\end{center}
\caption{\emph{{\bf Dependence of $dA/dt$ on $w$.} {\bf Left.} For $f=0.001$ and different values for $g$. {\bf B.} For fixed $g=0.001$, and different values of $f$. }} 
\label{symmetric_dA_dt}
\end{figure}

While the quantitative implementation is speculative, below we give a simplified possible shape for $dA/dt$ that incorporates these details, with a positive peak at $w_0$, decreasing to two negative dips, then slowly returning asymptotically to zero.  For now, we consider, for simplicity, that the deviations towards the high and low temperature range have symmetric effects on $A$:

\begin{equation}
\frac{dA}{dt} = \left[ 1 - g(w - w_0)^2 \right] e^{-f(w - w_0)^2}
\end{equation}

\noindent where $f$ and $g$ are sensitivity parameters which tune the width and steepness of the graph. Figure~\ref{symmetric_dA_dt} shows the graph of $dA/dt$ plotted with respect to temperature, $w$, based on different values for the parameters, $f$ and $g$. 

While the shape of the graph remains qualitatively the same, the function is highly sensitive quantitatively to small perturbations in either parameter.  Since these parameters influence the dependence of Carbon emissions on temperature, they encompass the effects of human behavior on this dependence. When interpreting our model, we view fine tuning of these parameters as corresponding adjustments in the human response to temperature changes. For a fixed $f$, changing $g$ affects primarily the negative critical values; that is, smaller values of $g$ lead to the graph reaching much shallower negative dips, without changing the position of the relaxation time. We interpret a decrease in $g$ as an increase in the rate of change of $A$ around the critical temperatures in areas regarding these negative dips. For a fixed $g$, changing $f$ affects not only the negative critical values, but also the positions of the critical points, and the relaxation time, with imperceptible changes, however, on the graph in the positive range. We interpret these as changes as different possibilities of extrema that are possible for $\frac{dA}{dt}$.

\section{Numerical simulations}
\label{numerics}

The first numerical difficulty we had to overpass was inherited from the original two-dimensional fast-slow system. The huge difference of almost four orders of magnitude between the time constants $R$ and $1/\varepsilon$ is unpleasant for both numerical and visualization purposes. Hence, before using any integration methods in our analysis, we performed a change of scale (e.i., a change in units of our time constants). Using a millennium ($10^3$ years $\sim 3.16 \times 10^{10}$ seconds) as our time unit, the original values of our time constants $R=4 \cdot 10^9$, $\Omega=1.5 \cdot 10^{11}$, $\varepsilon=3.9 \cdot 10^{-13}$ become $R=0.1266$, $\Omega = 0.474$, $\varepsilon=0.01264$. All simulations are performed with these transformed values, hence the results will be expressed on a time scale measured in thousands of years.

To simulate the asymptotic behavior of our system, we used Matcont continuation algorithms for finding equilibria, extending equilibrium curves with respect to the sensitivity parameters $f$ and $g$ and detecting bifurcations of codimension 1 along these curves. When using Matcont continuation algorithms, one needs to proceed with caution, since small variations in integration step or computation precisions have effects on the computation. Too large a step may lead to a variety of spurious results: e.g., and excessively coarse approximation may omit precisely the transition points one is searching for. On the other hand, too small a step may be too computationally expensive to permit a comprehensive search for bifurcations in reasonable time. Part of the difficulty of efficiently using the softwear for numerical computations which are not supported by analytical results is finding the optimal integration parameters.

\begin{figure}[h!]
\begin{center}
\includegraphics[width=0.95\textwidth]{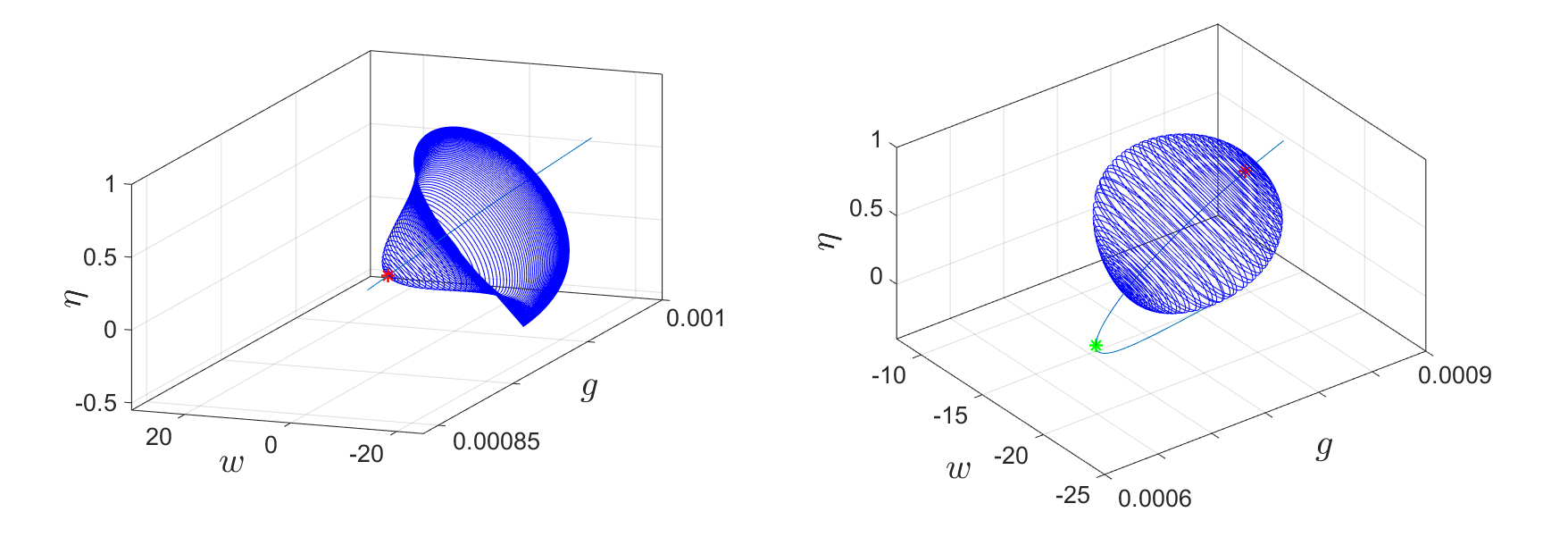}
\end{center}
\label{temp_Hopf bifurcations}
\caption{\emph{{\bf Subcritical Hopf bifurcation with respect to the parameter $g$.} {\bf Left.} The plot shows the equilibrium curve with respect to $g$, for fixed $f=0.001$. The Hopf point (which occurs approximately at $g=0.000854$  is marked on the curve by a red star; we also illustrate the evolution of the unstable cycle sprouting from the Hopf point, as we allow $g$ and the period to change. {\bf Right.} The same equilibrium curve and Hopf point are shown; the unstable cycle is extended instead with respect to both $g$ and $f$ simultaneoulsy. The wider coordinate window also includes in this case a limit point occurring approximately at $g=0.00066$ on the equilibrium curve (shown as a green star). All other system parameters were fixed for this simulation to:  $B=1.9$, $C=3.04$, $\alpha_w =0.32$, $\alpha_s=0.62$, $s_2=-0.482$, $w_0=20$, $T_c=-10$, $R=0.1266$, $\Omega = 0.474$, $\varepsilon = 0.01264$, $a=20$.}}
\label{fixed_f}
\end{figure}

\begin{figure}[h!]
\begin{center}
\includegraphics[width=0.55\textwidth]{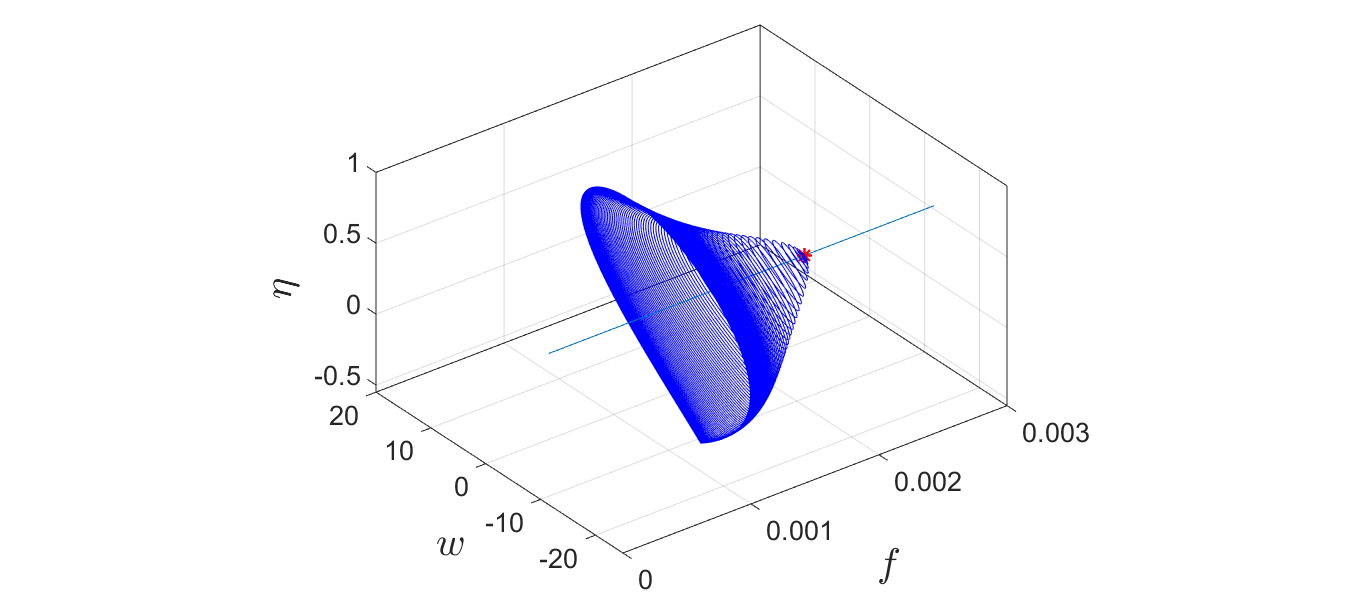}
\includegraphics[width=0.4\textwidth]{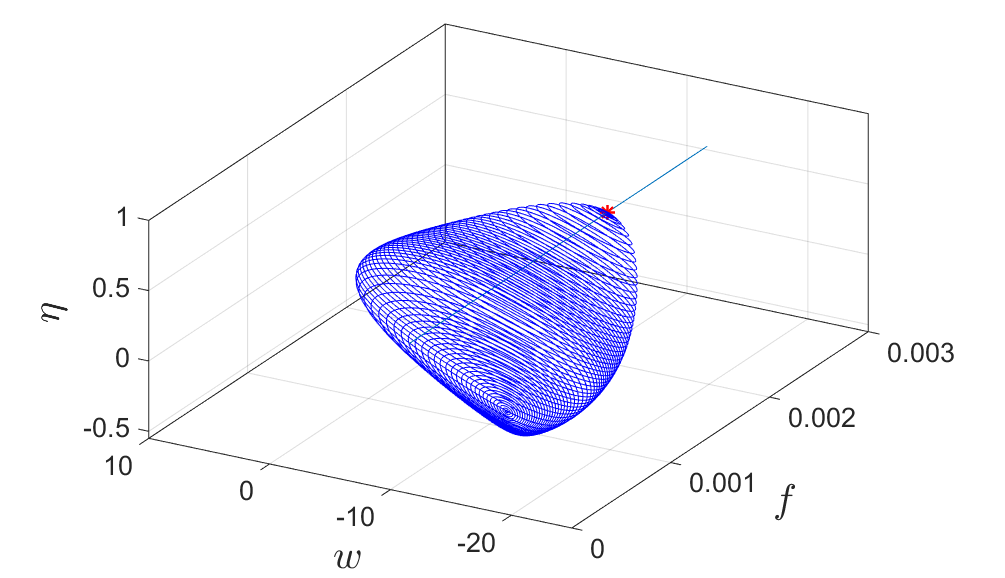}
\end{center}
\label{temp_Hopf bifurcations}
\caption{\emph{{\bf Supercritical Hopf bifurcation with respect to the parameter $f$.} The plot shows the equilibrium curve as $f$ changes in the corresponding interval, for fixed $g=0.001$.  The Hopf point (which occurs approximately at $f=0.002$  is marked on the curve by a red star; we illustrate the evolution of the stable cycle sprouting from the Hopf point, as we allow $f$ and the period to change. {\bf Right.} The same equilibrium curve and Hopf point are shown; the stable cycle is extended with respect to both $f$ and $g$ simultaneoulsy. All other system parameters were fixed for this simulation to:  $B=1.9$, $C=3.04$, $\alpha_w =0.32$, $\alpha_s=0.62$, $s_2=-0.482$, $w_0=20$, $T_c=-10$, $R=0.1266$, $\Omega = 0.474$, $\varepsilon = 0.01264$, $a=20$.}}
\label{fixed_g}
\end{figure}

The results were generally promising, and our simulations detected rich behavior along equilibrium curves, among which saddle node and subcritical Hopf bifurcations, implying the presence of limit cycles (see Figures~\ref{fixed_f} and~\ref{fixed_g}). But at a closer inspection, we found some aspects of the simulations to need additional justification and tuning in order to be representative of plausible behavior. For example: both our parameters $g$ and $f$ tune the sensitivity of $A$ to temperature -- for a fixed $f$, increasing $g$ exacerbates the negative effects of $w$ on $A$, and so does increasing $f$ when keeping $g$ fixed.

However, unlike one may expect at a first glance, the simulated effects of increasing the parameters $f$ and $g$ were very different. Increasing $g$ for fixed $f$ pushed the system through a Hopf bifurcation and changed the stability of the equilibrium from repelling to attracting spiral coexisting with an unstable cycle (Figure~\ref{fixed_f}). This was surprising, since one rather expects that, by increasing the sensitivity to CO$_2$, the system would become less stable. On the other hand, the effect was the opposite when decreasing $f$ for fixed $g$, and the equilibrium undergoes a Hopf bifurcation and becomes stable. These differences may be due to the fact that $g$ affects the $dA/dt$ curve along the whole effective range of temperatures, and $f$ more focused in the center of the domain, hence they may affect differently dynamics with our phase and parameter ranges.

Altogether, we realized that our construction needed more clarification, and a more qualitatively accurate set of sensitivity parameters to describe the dependence of $dA/dt$ on $w$. In Section~\ref{vapor_section}, we present a refinement of our model.

\subsection{Introducing water vapor}
\label{vapor_section}

\begin{figure}[h!]
\begin{center}
\includegraphics[width=0.48\textwidth]{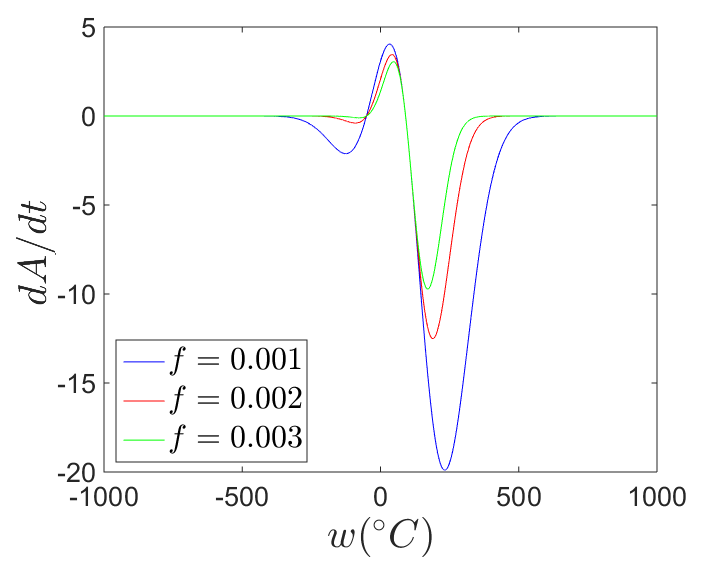}
\includegraphics[width=0.48\textwidth]{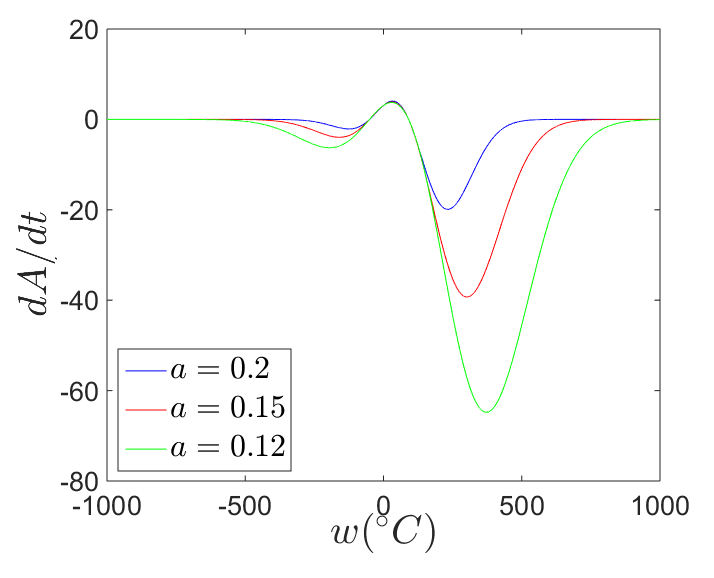}
\end{center}
\caption{\emph{{\bf Graphs of $dA/dt$ (with water vapor included) with respect to $w$.} On the left we see different values of $f$ with fixed $g = 0.001 $ and $\alpha = 0.2 $. On the right we see different values of $a$ with fixed $f = 0.001$ and $g = 0.001$}}
\label{vapor_graphs}
\end{figure}

One major effect which was omitted in our first extension was the difference between physical and chemical behavior of elements at high and low temperature extremes. Unlike low temperatures, high surface temperature raise the amounts of vapors present in the atmosphere, which, aside from CO$_2$, may contribute substantially to the greenhouse effect (hence to the behavior of our $A$ in the model). Although the subject is still controversial in the climatology literature, some studies claim in fact that the largest greenhouse effect is in fact that of water vapors in the atmosphere (rather than CO$_2$). To investigate the possible consequences of these ideas, we decided to introduce the effect of evaporation in the model, and study how this changed the model predictions. We increased the negative effect on A at high temperatures, where the water is expected to become vapor and raise up in the atmosphere, contributing to the greenhouse effect, so that

\begin{equation}
\frac{dA}{dt} = \left[M - g(w - w_0)^2 \right]e^{-f(\alpha w - w_0)^2}
\label{vapos_bias}
\end{equation}

\noindent where $\alpha$ governs the asymmetry in the graph due to vapor bias at high temperatures: more water vapor leads to less OLR and the negative rate of change will be larger in a range of very high temperatures where evaporation occurs. For temperatures even beyond that range, the graph decays to zero, as before. Figure~\ref{vapor_graphs} shows a few examples of curves $dA/dt$ for a few representative values of $f$, $g$ and $a$, illustrating the effect that each has on the shape of the graph.

In our simulations, we again observed the effects of changing the sensitivity parameters $f$ and $g$, for various values of $a$ chosen to that the evaporation range (and the corresponding local minimum) remain within plausible bounds. The effects are now closer to what one would expect from a system of such complexity.

First, aside from stable equilibria, the system exhibits stable limit cycles for certain parameter ranges, and possibly other stable invariant sets, which we have not yet investigated. A locally stable cycle is very significant, since it can stabilize long-term behavior (for initial conditions in a certain attraction basin) to oscillations between low and high temperatures/iceline states. These could be fast oscillations (if the cycle has small period), or long oscillations, on the time-scale of ice ages (if the period is large). 

\begin{figure}[h!]
\begin{center}
\includegraphics[width=0.7\textwidth]{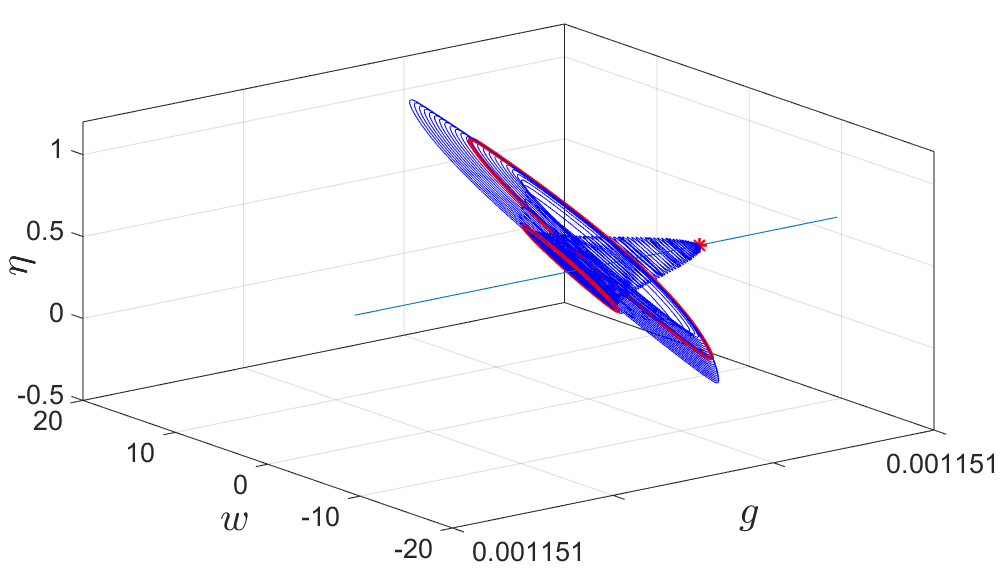}
\end{center}
\label{temp_Fold bifurcations}
\caption{\emph{{\bf Sequence of Hopf-fold bifurcation with respect to the parameter $g$.} The equilibrium curve with respect to $g$ has a supercritical Hopf bifurcation at aproximately $g=0.0011512$, with the birth of a stable cycle. Following the evolution of this cycle (by changing $g$ and the period), we found two consecutive fold bifurcations, the first at $g \sim 0.00115118$, and the second at $g \sim 0.001151187$.  All other system parameters were fixed for this simulation to:  $B=1.9$, $C=3.04$, $\alpha_w =0.32$, $\alpha_s=0.62$, $s_2=-0.482$, $w_0=20$, $T_c=-10$, $R=0.1266$, $\Omega = 0.474$, $\varepsilon = 0.01264$, $M=$, $\alpha=0.008$, $f=0.002$.}}
\label{fold}
\end{figure}

We found that, when either $f$ or $g$ are decreased, the system may switch from a stable equilibrium to a stable cycle, with increasingly large period as the parameter changes. However, these stable cycles themselves may only survive within a relatively small parameter range. For example, Figure~\ref{fold} shows how decreasing $g$ transitions a stable equilibrium, though a supercritical Hopf bifurcation (marked as a red star) to an unstable equilibrium, with the birth of a stable cycle. The cycle itself bifurcates through a fold bifurcation, giving birth to another, unstable cycle (the left red cycle marks the fold bifurcation, where these two cycles coincide). When continuing to extend the unstable cycle, it also crosses a fold bifurcation (right red cycle), producing a a larger, stable cycle. For a very small interval of $g$, the system has two stable cycles (bistability), separated by an unstable cycle. 

Bistability is a very plausible and desired features for natural function, allowing the system to converge to two different outcomes depending on the initial conditions. We suspect that complex systems put a lot of effort into functioning in a critical range (close to a bifurcation state) where this type of transitions are facilitated, to allow the system to respond swiftly to change.

\section{Discussion}

Our paper shows how small changes in the parameters of our climatic system may lead to dramatic variations in its long term dynamics. We focused in particular on parameters that described in our model the impact on the climate of known human activity (interpreted from the way in which they affected the level of green house gases produced). We may conclude that even subtle changes in our patterns (e.g., exercising moderation -- or alternative, non-polluting resources -- when using heating/cooling according to seasonal variations, or in response to major cold or hot waves) may lead to a better climatic prognosis. This may be hard, since the climate has already started to exhibit wide climate (e.g., temperature, precipitation) swings over short periods of time, which may induce people to use more gass-producing resources in order to maintain their current life style. Hence we find it crucial to point out how important it is to try to stick with more conservative patterns of traditional energy consumption.

Climate change is an important subject is the field of mathematical modeling. The Budyko system had been studied a lot over the past few decades, and we believe that there is much more to be done. It is an over simplified model of Earth's climate, but studying it can help us understand the essential behavior of our climate, and how our patterns affect it.

Altogether, it is extremely important to understand, mathematically and philosophically, how the term of ``global warming'' (defined as the effects of greenhouse gases on the Earth's climate) does not necessarily imply only a steady increase in the planet's annual average temperature (concept still unfortunately used by some scientist to deny these effects, see e.g. \url{http://www.climatedepot.com}), but can in fact refer to much more complex phenomena. Wild cycling between extreme phenomena such as hot and cold periods (from hotter summer and colder winters, to hotter decades and colder decades, etc) may lead to seemingly steady average behavior, without describing the essence of the climatic dynamics.

There is a great amount of work yet to be done on the model, on aspects such as: (1) further refine the model of $dA/dt$, fine tune parameters to integrate the behavior within realistic functional ranges and, if possible, validate it with empirical measurements; (2) interprete more carefully the contribution of each parameter, trying to segregate the aspects that can be controlled by human behavior from the aspects that cannot be altered, but only understood.

\bibliographystyle{plain}
\bibliography{Luke_references}

\end{document}